\documentclass[a4paper,11pt]{article}
\pdfoutput=1 

\usepackage{jcappub} 

\usepackage[T1]{fontenc} 

\title{A parametrisation of modified gravity on nonlinear cosmological scales}

\author[a]{Lucas Lombriser}

\affiliation[a]{Institute for Astronomy, University of Edinburgh, Royal Observatory, Blackford Hill, Edinburgh, EH9~3HJ, U.K.}

\emailAdd{llo@roe.ac.uk}

\abstract{
Viable modifications of gravity on cosmological scales predominantly rely on screening mechanisms to recover Einstein's Theory of General Relativity in the Solar System, where it has been well tested.
A parametrisation of the effects of such modifications in the spherical collapse model is presented here for the use of modelling the modified nonlinear cosmological structure.
The formalism allows an embedding of the different screening mechanisms operating in scalar-tensor theories through large values of the gravitational potential or its first or second derivatives as well as of linear suppression effects or more general transitions between modified and Einstein gravity limits.
Each screening or suppression mechanism is parametrised by a time, mass, and environment dependent screening scale, an effective modified gravitational coupling in the fully unscreened limit that can be matched to linear theory, the exponent of a power-law radial profile of the screened coupling, determined by derivatives, symmetries, and potentials in the scalar field equation, and an interpolation rate between the screened and unscreened limits.
Along with generalised perturbative methods, the parametrisation may be used to formulate a nonlinear extension to the linear parametrised post-Friedmannian framework to enable generalised tests of gravity with the wealth of observations from the nonlinear cosmological regime.
}

\newcommand{\bq}{\begin{equation}}
\newcommand{\eq}{\end{equation}}
\newcommand{\bqa}{\begin{eqnarray}}
\newcommand{\eqa}{\end{eqnarray}}

\newcommand{\rmd}{\ensuremath{\mathrm{d}}}

\newcommand{\Om}{\Omega_{\rm m}}
\newcommand{\Olam}{\Omega_{\Lambda}}
\newcommand{\rhom}{\rho_{\rm m}}
\newcommand{\rhomb}{\bar{\rho}_{\rm m}}
\newcommand{\drhom}{\delta\rhom}
\newcommand{\scal}{\varphi}
\newcommand{\bscal}{\bar{\scal}}
\newcommand{\dscal}{\delta\scal}

\newcommand{\Geff}{G_{\rm eff}}

\newcommand{\RTH}{r_{\rm th}}
\newcommand{\yhal}{y_{\rm h}}
\newcommand{\yenv}{y_{\rm env}}
\newcommand{\deltac}{\delta_{\rm c}}

\begin{document}
\maketitle
\flushbottom

\section{Introduction}

The length scales involved when performing cosmological tests of gravity compare to the more conventional probes in the Solar System like the extrapolation in orders of magnitude from the scale of human experience to the scale of an atomic nucleus.
It is therefore well worth inferring independent constraints on the gravitational interactions in the cosmological regime.
The past decade has seen a steep growth in such cosmological
experiments and much progress has been made in
measuring
the gravitational force
over the vast scales of our Universe (see, e.g., Ref.~\cite{koyama:15} for a review).
In order to comply with the stringent bounds set by the Solar-System tests, where Einstein's Theory of General Relativity (GR) has been verified~\cite{will:05}, the modified gravity models under scrutiny predominantly rely on screening mechanisms~\cite{vainshtein:72,khoury:03a,babichev:09,hinterbichler:10,brax:10,lombriser:14b,mcmanus:16} that suppress the modifications in high-density regions while allowing significant modifications at the larger, cosmological scales of lower density.
While the different screening mechanisms vary in their attributes, featuring dependencies on the mass or density of a system as well as its morphology and environment, they all exhibit a transition between a large-scale modification and a small-scale recovery of GR at some characteristic screening scale.

Modifications of gravity in the cosmological context have been motivated for a variety of purposes.
Generally, gravitational physics is not understood in the ultraviolet and a more fundamental theory possibly unifying GR with the
Standard Model interactions
may give rise to a remnant extra degree of freedom in the infrared, which could affect the late-time expansion and large-scale structure of the Universe.
Hence, traditionally, modifications of gravity on cosmological scales
have also been considered as an explanation for the observed cosmic acceleration, alternative to the inexplicably small cosmological constant $\Lambda$ of the concordance $\Lambda$ Cold Dark Matter ($\Lambda$CDM) model or
to dark energy (see Refs.~\cite{koyama:15,clifton:11,joyce:16} for reviews).
It can, however, be shown that the most general modified gravity theory involving a single extra scalar degree of freedom with second-order equations of motion (Ref.~\cite{horndeski:74}), which is the framework embedding most of the alternatives proposed, cannot yield an observationally consistent self-acceleration that is genuinely different from the contribution of dark energy or a cosmological constant, provided a cosmological propagation of gravitational waves at the speed of light~\cite{lombriser:15c,lombriser:16}.
This conclusion does likely not apply more broadly to theories with other or multiple extra degrees of freedom.
However, it should also be noted that more complex theories generally introduce more freedom, which may ultimately produce degeneracies that cannot be broken by
observations (see, e.g., Ref.~\cite{lombriser:14b}).

Regardless of the cosmic acceleration problem, it is worthwhile testing for a non-minimal coupling of an extra degree of freedom to the metric or the different matter species.
Instead of a universal coupling, however, one may also consider the separation of couplings between the different matter components.
In this case, cosmological constraints, mostly testing the coupling to dark matter, can be regarded as independent of the local and astrophysical bounds that predominantly rely on interactions with baryons.
The former may be viewed as an interaction between dark energy and dark matter but if a cosmological constant serves as the dominant driver of cosmic acceleration, the new degree of freedom may rather be interpreted as the force carrier of an interaction between the dark matter particles.
It should be emphasised that many simulations and observational tests of modified gravity on cosmological scales neglect the influence of baryons and, hence, may to some extent be reinterpreted within the context of those alternative scenarios.
In this respect, screening may still be required in some cases to allow for significant interactions of dark matter particles at large scales that are compatible, for instance, with the observed cluster abundance~\cite{schmidt:09c,lombriser:10,cataneo:14,liu:16,peirone:16}.
Finally, note that the extra degree of freedom may serve as a dark matter candidate itself (see, e.g.,~\cite{marsh:15,urena-lopez:15}).

Given the plethora of modified gravity or dark sector interactions possible based on the prospects of new fields, the development of a more systematic approach to explore their cosmological implications has been and continues to be of great interest.
The generalised framework for the resulting modifications in the formation of structure should provide a consistent description on large and small scales.
For this purpose, parametrised post-Friedmannian (PPF) formalisms~\cite{uzan:06,caldwell:07,zhang:07,amendola:07,hu:07b,bertschinger:08,daniel:10,lombriser:11a,dossett:11,hojjati:11,battye:12,baker:12,lombriser:13a,silvestri:13} have been developed, inspired by the parametrised post-Newtonian expansion in a low-energy static limit~\cite{will:05}, or similarly an effective field theory (EFT) of dark energy~\cite{creminelli:08,park:10,gubitosi:12,bloomfield:12,tsujikawa:14,bellini:14,gleyzes:14b}
(also see applications in Refs.~\cite{hu.b:13,raveri:14,lombriser:14b,lombriser:15b,valkenburg:15,lombriser:15c,linder:15,lagos:16,zumalacarregui:16,Pogosian:16}).
These
unify the computation of the evolution of the spatially homogeneous and isotropic cosmological background and the linear perturbations around it for a large class of modified gravity and dark energy scenarios, covering structures from the Hubble scale to a few tens of Mpc.
Eventually, however, linear theory fails in describing the cosmological structure below these scales.
Together with the screening effects, this severely complicates performing consistent tests of gravity. The more complex nonlinear structures observed, for instance, in the abundance of clusters, environmental and symmetry dependencies, substructure, and the cosmic web, however, conceal a great amount of information about our Universe and the physical processes at work (see, e.g.,~\cite{llinares:12,shim:13,hellwing:13,burrage:14,lombriser:15a,shi:15,falck:15,thomas:15,he:16,ivarsen:16,stark:16,baldi:10}).
In particular, in theories introducing a mass scale larger than order ${\rm Mpc}^{-1}$ like chameleon gravity~\cite{khoury:03a}, gravitational forces are only enhanced in the nonlinear cosmological structure, rendering it a vital regime for testing gravity.
Notably, screening effects, along with other nonlinear processes, can also prevent strong anomalies from manifesting in measurements of the averaged structure, which limits the constraints that can be inferred unless, for instance, employing statistical techniques devised to unscreen those~\cite{lombriser:15a}.
For the study of the nonlinear regime of structure formation and the suppression effects in the gravitational modifications, $N$-body simulations provide an indispensable tool (see Ref.~\cite{winther:15} for a review).
In general, these simulations are, however, computationally considerably more expensive than their Newtonian counterparts and do not cover the plethora of gravity and dark sector models proposed.
Hence, for comprehensive comparisons of theory to observations, allowing a full exploration of the cosmological and model parameter spaces involved, more efficient and generalised modelling techniques are required.

A nonlinear extension of the linear PPF formalism is therefore highly desirable to correctly interpret and exploit the rich cosmological survey data available below a few tens of Mpc.
A variety of approaches
to this objective
have been
pursued,
employing phenomenological frameworks and fitting functions to simulations~\cite{hu:07b,zhao:10b,li.y:11,zhao:13,mead:16}, the rescaling~\cite{mead:14} or speed-up~\cite{winther:14,barreira:15b} of simulations, analytical and numerical approximations~\cite{schmidt:10,pourhasan:11,lombriser:12,terukina:12,terukina:13,gronke:15,hammami:16,sakstein:16}, or perturbation theory~\cite{koyama:09,brax:13,taruya:16,bose:16}.
Thereby, nonlinear PPF formalisms have been proposed based on interpolations between the modified and screened regions of the matter power spectrum~\cite{hu:07b} calibrated to simulations, or similarly using an effective variance~\cite{li.y:11}.
Of particular interest has been the study of modified gravity effects in the spherical collapse model and its applications~\cite{schmidt:08,schmidt:09b,borisov:11,li:11b,lombriser:11b,li:12,lam:12,clampitt:12,lombriser:13b,pace:13,lombriser:13c,barreira:14a,kopp:13,taddei:13,barreira:13,lombriser:14a,brax:14b,lam:14,gomez:15,achitouv:15,cataneo:16}.
Examples include $f(R)$ gravity~\cite{buchdahl:70,schmidt:08,borisov:11,lombriser:13b} or more general chameleon theories~\cite{khoury:03a,li:11b,lombriser:13c}, symmetron models~\cite{hinterbichler:10,taddei:13}, the Vainshtein mechanism in Dvali-Gabadadze-Porrati (DGP) braneworld gravity~\cite{vainshtein:72,dvali:00,schmidt:09b} or galileon models~\cite{nicolis:08,barreira:13}, k-mouflage models~\cite{babichev:09,brax:14b}, and unscreend, linear PPF modifications~\cite{gomez:15}.
Applications include, for instance, excursion set theory~\cite{li:11b,li:12,lam:12,clampitt:12,lombriser:13b,kopp:13} and the modelling of halo~\cite{schmidt:08,schmidt:09b,lombriser:13c,barreira:14a}
or void properties~\cite{clampitt:12,lam:14}.
For chameleon gravity, a review and comparison of the different methods can be found in Ref.~\cite{lombriser:14a}.
In general, spherical collapse computations have proven very useful in capturing the modified gravity effects in the nonlinear structure formation.

Motivated by the variety of screening mechanisms available to scalar-tensor theories, this paper introduces a
parametrisation of the modified gravitational forces that can act on spherical top-hat overdensities.
Applied to the spherical collapse model, it is then used to formulate a nonlinear extension to the linear PPF formalism that enables tests of generalised modifications with the deeply nonlinear cosmological structure.
The paper is organised as follows.
Sec.~\ref{sec:sphcollapse} briefly reviews the main aspects of the spherical collapse model in the context of modified gravity and introduces a parametrisation of the effective gravitational coupling.
A discussion of how the different screening mechanisms can be mapped onto this effective coupling is given in Sec.~\ref{sec:mapping}.
Sec.~\ref{sec:einsteinlimits} then describes how the scaling method of Ref.~\cite{mcmanus:16} can be used to embed more general transitions from a modified to an Einstein gravity limit.
Sec.~\ref{sec:PPF} is devoted to the implementation of these results in a nonlinear PPF formalism.
Finally, a conclusion of this programme is presented in Sec.~\ref{sec:conclusions}.

\section{Generalised spherical collapse model} \label{sec:sphcollapse}

The nonlinear cosmological structure formation can be studied with the spherical collapse model, where a dark
matter halo is approximated by a spherically symmetric top-hat overdensity and evolved according to the nonlinear continuity and Euler equations from an initial era to the time of its collapse.
Modified gravity effects can be incorporated in these calculations by accounting for an effective modification of the gravitational coupling that enters through the gravitational potential in the momentum conservation equation.
Sec.~\ref{sec:sphcoll} provides a brief derivation of the spherical collapse equation, where the effect of modifying gravity is presented in Sec.~\ref{sec:MGsphcoll}.
Sec.~\ref{sec:Geff} introduces a parametrisation for screening mechanisms and other suppression effects on the effective modified gravitational coupling.
%

\subsection{Collapse of a spherical top-hat density} \label{sec:sphcoll}

The nonlinear continuity and Euler equations of a metric theory of gravity in comoving spatial coordinates for a pressureless non-relativistic matter fluid  are given by~\cite{peebles:80,schmidt:08,pace:10}
\bqa
 \dot{\delta} + \frac{1}{a}\nabla\cdot(1+\delta)\mathbf{v} & = & 0 \,, \\
 \dot{\mathbf{v}} + \frac{1}{a}\left(\mathbf{v}\cdot\nabla\right)\mathbf{v} + H \mathbf{v} & = & -\frac{1}{a}\nabla\Psi \,,
\eqa
where dots represent derivatives with respect to physical time, $\delta\equiv\drhom/\rhomb$, $\Psi\equiv\delta g_{00}/(2g_{00})$ denotes the gravitational potential, and $H\equiv\dot{a}/a$ is the Hubble parameter.
The two equations combine to
\bq
 \ddot{\delta} + 2H\dot{\delta} - \frac{1}{a^2}\nabla_i\nabla_j(1+\delta)v^iv^j = \frac{1}{a^2}\nabla_i(1+\delta)\nabla^i\Psi \,.
\eq
For the velocity term, one may adopt the approximation of a spherical top-hat density with $\mathbf{v}=A(t)\mathbf{r}$, which can be combined with the continuity equation to infer
\bq
 \frac{1}{a^2}\nabla_i\nabla_jv^iv^j=\frac{4}{3}\frac{\dot{\delta}^2}{(1+\delta)^2} \,.
\eq
Hence, the spherical collapse equation becomes
\bq
 \ddot{\delta} + 2H\dot{\delta} - \frac{4}{3}\frac{\dot{\delta}^2}{(1+\delta)^2} = \frac{1+\delta}{a^2}\nabla^2\Psi \,. \label{eq:sphcoll}
\eq
Mass conservation implies constant $M=(4\pi/3)\rhomb(1+\delta)\zeta^3$, where $\zeta(a)$ denotes the physical top-hat radius at $a$.
In combination with Eq.~(\ref{eq:sphcoll}) this yields
\bq
 \frac{\ddot{\zeta}}{\zeta} = H^2 + \dot{H} - \frac{1}{3a^2}\nabla^2\Psi \,, \label{eq:shellevol}
\eq
which describes the evolution of a spherical shell at the edge of the top hat.

\subsection{Spherical collapse in modified gravity} \label{sec:MGsphcoll}

The impact of a modification of gravity on the spherical collapse of the top hat can be captured by an effective modification of the Poisson equation entering Eq.~(\ref{eq:shellevol}) and may be parametrised as
\bq
 \nabla^2\Psi \equiv \frac{a^2}{2} \left(1+\frac{\Delta\Geff}{G}\right) \kappa^2 \drhom \,,
\eq
where $\kappa^2\equiv8\pi\,G$ with bare gravitational constant $G$, such that the spherical collapse equation becomes
\bq
 \frac{\ddot{\zeta}}{\zeta} = H^2 + \dot{H} - \frac{\kappa^2}{6} \left(1+\frac{\Delta\Geff}{G}\right) \drhom \,. \label{eq:MGsphcoll}
\eq
Note that the metric field equations for a modification of gravity can be rewritten as
\bq
 G^{\mu\nu} \equiv \kappa^2\left(T^{\mu\nu}+T^{\mu\nu}_{\rm eff}\right) \,, \label{eq:Tmunueff}
\eq
where $G^{\mu\nu}$ is the Einstein tensor and $T^{\mu\nu}_{\rm eff}$ denotes an effective energy-momentum tensor embodying all extra terms of the new field equation.
With this definition, the first two terms on the right-hand side of Eq.~(\ref{eq:MGsphcoll}) may also be written as
\bq
 H^2 + \dot{H} = -\frac{\kappa^2}{6} \left[ \rhomb + ( 1+3w_{\rm eff}) \bar{\rho}_{\rm eff} \right] \,,
\eq
which follows from using the Friedmann equations obtained with Eq.~(\ref{eq:Tmunueff}).

To solve Eq.~(\ref{eq:MGsphcoll}), let $\RTH$ denote the comoving top-hat radius with $\zeta(a_i)=a_i\RTH$ at an initial scale factor $a_i\ll1$.
We shall further define $y\equiv \zeta/(a\,\RTH)$, where from mass conservation, $\rhomb a^3 \RTH^3 = \rhom \zeta^3$, it follows that $\rhom/\rhomb=y^{-3}$.
With these definitions Eq.~(\ref{eq:MGsphcoll}) becomes
\bq
 y'' + \left(2+\frac{H'}{H}\right) y' + \frac{1}{2}\Om(a)\left(1+\frac{\Delta\Geff}{G}\right) \left(y^{-3}-1\right) y = 0 \,, \label{eq:ydiff}
\eq
where $\Om(a)\equiv\kappa^2\rhomb/(3H^2)$ and
primes denote derivatives with respect to $\ln a$.
One can solve this differential equation setting the initial conditions at $a_i\ll1$ in the matter-dominated regime, $y_i \equiv y(a_i) = 1 - \delta_i/3$ and $y_i' = - \delta_i/3$.

The spherical collapse density $\deltac(z)$ is defined by the extrapolation of the initial overdensity $\delta_i$ that yields collapse in Eq.~(\ref{eq:ydiff}) at a given redshift $z$ using the linear growth factor $D/D_i\equiv\delta_{\rm lin}/\delta_i$, which is obtained from solving the linearisation of Eq.~(\ref{eq:MGsphcoll}).
In modified gravity models, $D$ can be both time- and scale-dependent but we may adopt an effective linear collapse density, defined by the extrapolation of $\delta_i$ with a GR growth function $D_{\rm GR}$, obtained from setting $\Delta\Geff=0$, which is independent of scale.

It should be noted that Birkhoff's theorem can be violated in modified gravity theories, causing shell crossing and a departure of the evolving overdensity from its initial top-hat profile~\cite{borisov:11,kopp:13}.
The top-hat evolution adopted here, however, has been shown to provide a good approximation if in these cases additionally accounting for the evolution of the environmental density surrounding the top hat and its impact on the effective gravitational coupling  $\Geff$~\cite{li:11b,li:12,lombriser:13b,lombriser:13c}.
In such scenarios, e.g., in chameleon gravity models, a coupled differential equation analogous to Eq.~(\ref{eq:ydiff}) for the dimensionless radius of the interior overdensity, $y\rightarrow\yhal$, will have to be solved simultaneously for $\yenv$.
%

\subsection{Effective screening} \label{sec:Geff}

Viable modified gravity theories introducing deviations from GR at large scales need to recover Einstein gravity in high-density regions in order to comply with the tight constraints inferred from Solar-System tests, which reflects in $\Delta\Geff/G\rightarrow0$.
This is predominantly achieved through nonlinear screening mechanisms introducing a characteristic scale where modified gravity transitions to GR (see~\cite{koyama:15,joyce:14,joyce:16} for reviews).
Similarly, modified gravitational forces can be limited to a finite range associated with the mass of the extra degree of freedom introduced.
This leads to a Yukawa suppression of the extra force beyond the Compton wavelength.
Ref.~\cite{lombriser:14b} pointed out that modified gravity theories can also give rise to a contrary effect, where gravity is modified on large scales but GR is recovered on small scales due to a linear shielding mechanism opposite to a Yukawa suppression.

This section proposes a parametrisation of the different effects of screening or linear suppression mechanisms on the modified spherical collapse through $\Delta\Geff/G$.
It will then be shown in Sec.~\ref{sec:mapping} how the different mechanisms can be mapped onto this parametrisation.
For this purpose, consider an effective gravitational coupling described by
\bq
 \frac{G_{eff}}{G} = A + \sum_i^{N_0} B_{i} \prod_j^{N_i} b_{ij} \left(\frac{r}{r_{0ij}}\right)^{a_{ij}} \left\{\left[1+\left(\frac{r_{0ij}}{r}\right)^{a_{ij}}\right]^{1/b_{ij}} - 1 \right\} \,, \label{eq:effscr}
\eq
where $i,j$ are positive integers.
Eq.~(\ref{eq:effscr}) simply defines a combination of interpolations between regimes of different radial dependence, where the particular form of the bracketed terms can be motivated by the Vainshtein mechanism (see Sec.~\ref{sec:vainshteinscr}).
The parameter $A$ describes the modification of the gravitational coupling in the fully screened limit, which could be different from unity. For one summand ($N_0=1$) and factor ($N_1=1$) in Eq.~(\ref{eq:effscr}), $B$ is the effective enhancement in the fully unscreened limit, $r_0$ is the screening scale, and $a$ determines the radial dependence of the screened solution (not to be confused with the scale factor of the metric) along with $b$ that defines an interpolation rate between the screened and unscreened limits.
The product takes into account multiple screening or suppression effects (with number of factors $N_i$ for each summand), e.g., a Yukawa suppression on large scales in addition to a chameleon screening on small scales.
The sum allows to describe a change of the screened gravitational coupling from $\Geff/G=A$ to a different value overlapping other screening mechanisms (with number of summands $N_0$).
In GR, $\Geff=G$ and we shall define $\Delta\Geff/G\equiv\Geff/G-1$.

As we will see in Sec.~\ref{sec:vainshteinscr}, for instance, we only require 4 parameters to describe the Vainshtein mechanism in the DGP model: the amplitude of the unscreened modification $B$, the screening scale $r_0$, the radial dependence in the screened limit $a(b-1)/b$, and the rate of interpolation $b$.
In chameleon models we need to define 7 parameters to include the Yukawa suppression on top of the screening effect (Sec.~\ref{sec:chameleonscr}).

Generally, for a term like
\bq
 \frac{\Delta G_{eff}}{G} \sim b \left(\frac{r}{r_0}\right)^a \left\{\left[1+\left(\frac{r_0}{r}\right)^a\right]^{1/b} - 1 \right\} \,,
\eq
we have the limits
\bq
 \frac{\Delta G_{eff}}{G} \sim \left\{
 \begin{array}{ll}
  b \left( \frac{r}{r_0} \right)^{a(b-1)/b} \,, & {\rm for} \; (b>0) \bigwedge \left[(r \ll r_0, a>0) \bigvee (r \gg r_0, a<0)\right] \,, \\
 -b \left( \frac{r}{r_0} \right)^a \,, & {\rm for} \; (b<0) \bigwedge \left[(r \ll r_0, a>0) \bigvee (r \gg r_0, a<0) \right] \,, \\
  1 \,, &  {\rm for} \; (r \ll r_0, a<0) \bigvee (r \gg r_0, a>0) \,.
 \end{array}
 \right. \label{eq:limits}
\eq
Near the screening scale, one also gets to first order
\bq
 r \rightarrow r_0 \; : \; \frac{\Delta G_{eff}}{G} \sim (2^{1/b}-1)b + a \left[ 2^{-1+1/b}(2b-1) -b \right] \left( \frac{r}{r_0}-1 \right) + \mathcal{O}\left[\left( \frac{r}{r_0}-1 \right)^2\right]\,, \label{eq:screeningscaleeq}
\eq
where one can use the zeroth order to calibrate the interpolation rate $b$ for the different screening mechanisms (Sec.~\ref{sec:mapping}).

\begin{figure}
 \centering
  \resizebox{0.496\hsize}{!}{\includegraphics{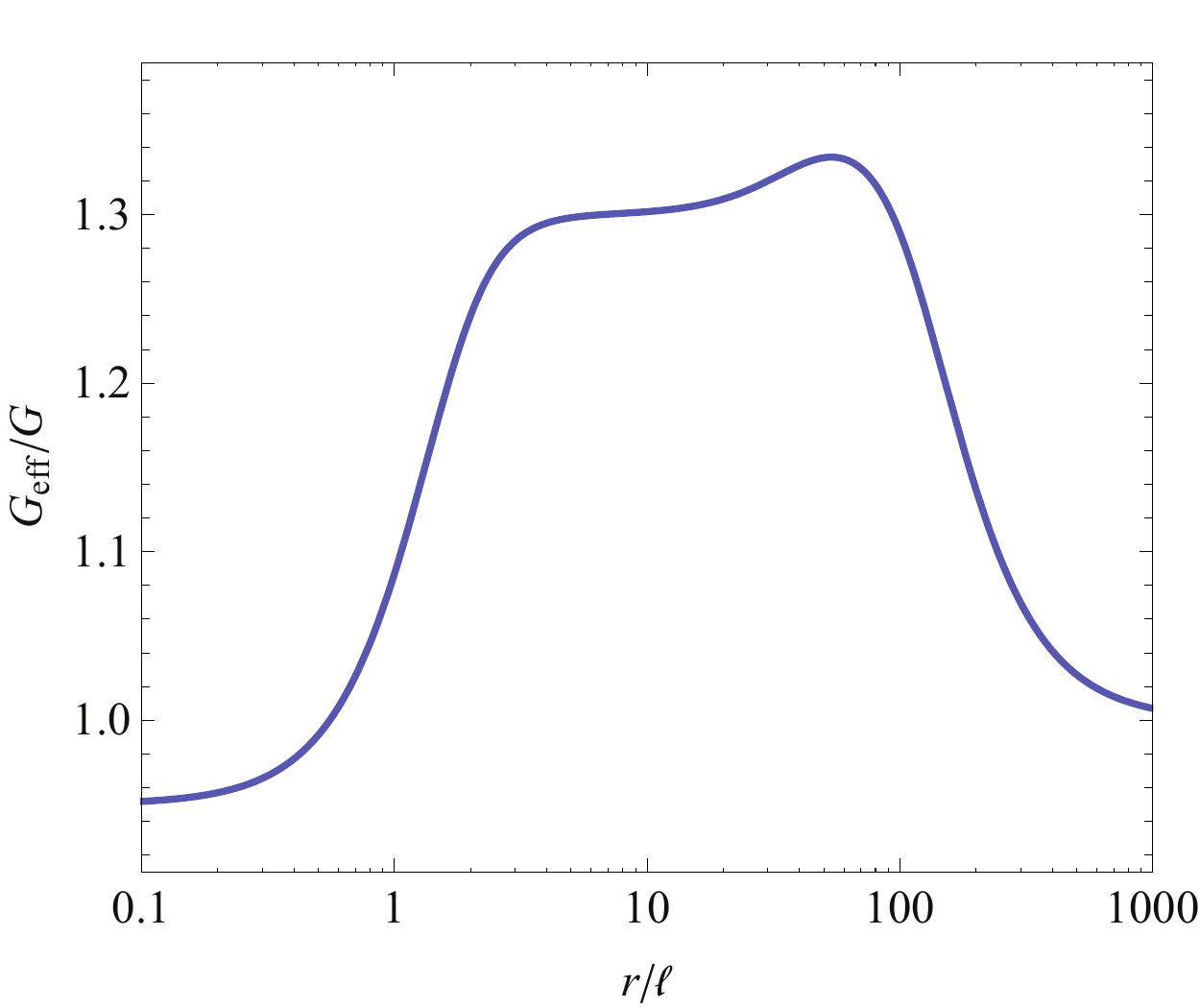}}
  \resizebox{0.496\hsize}{!}{\includegraphics{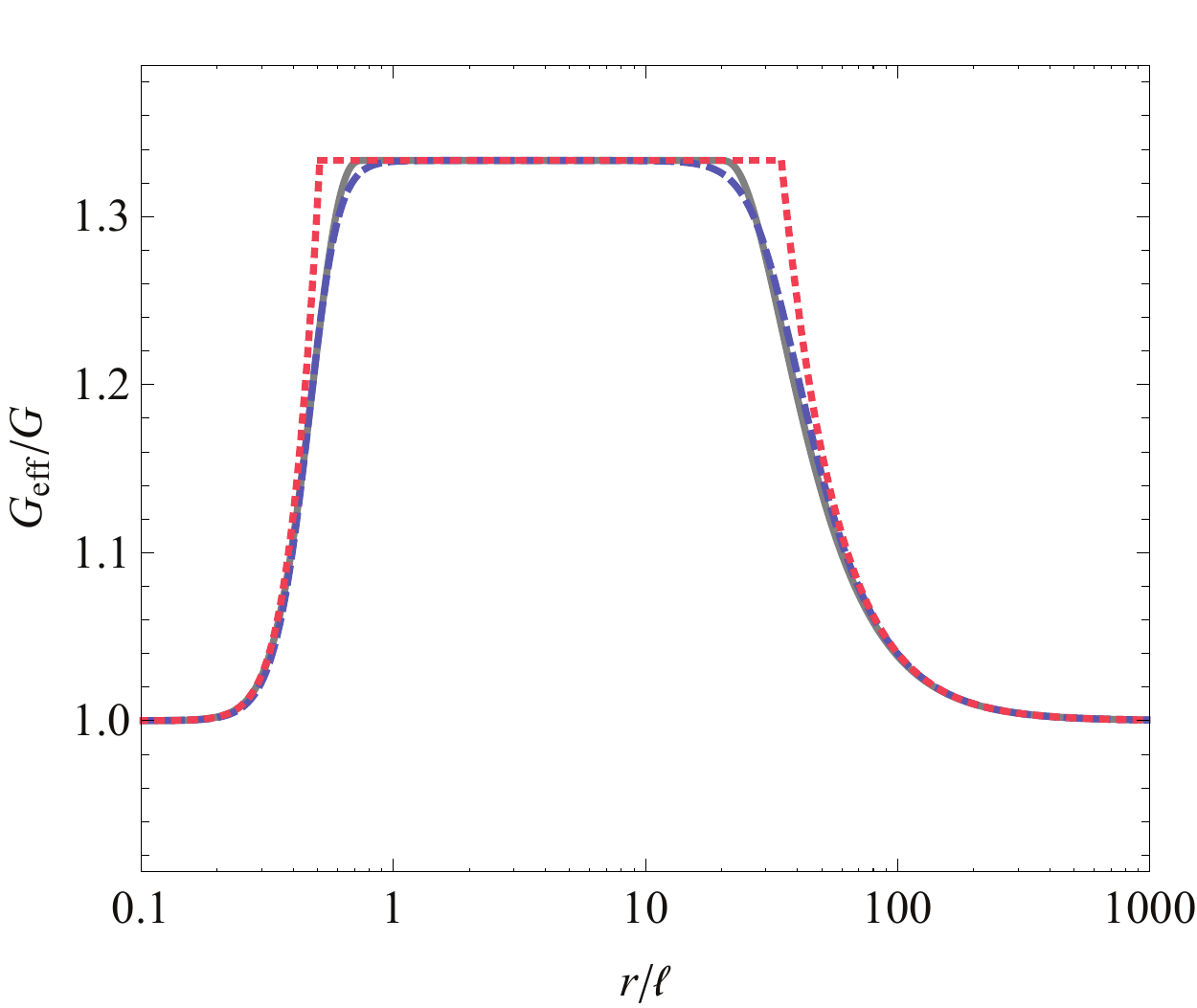}}
\caption{
Radial profile of the relative effective gravitational coupling. \textit{Left:}~Phenomenological example of a screening effect with ``overscreened'' centre, overlaying transition of the suppressed coupling, and Yukawa suppression, corresponding to the choices $(N_0,N_1,N_2)=(2,2,1)$, $(A,B_1,B_2)=(0.95,0.3,0.05)$, $(r_{011}\ell^{-1},a_{11},b_{11})=(2,4,2)$,
$(r_{012}\ell^{-1},a_{12},b_{12})=(100,-4,2)$, and $(r_{021}\ell^{-1},a_{21},b_{21})=(50,4,3)$ with length scale $\ell$.
\textit{Right:}~Chameleon screening effect. The gray solid curve shows the thin-shell approximation discussed in Sec.~\ref{sec:chameleonscr}. The red dotted curve corresponds to the linearised argument in the minimum function of Eq.~(\ref{eq:Geffchameleon}), which is often adopted as further approximation in the thin-shell approach (see discussion in Ref.~\cite{lombriser:13b}). The blue dashed line shows the parametrised effect with Eq.~(\ref{eq:effscr}). 
\label{fig:one}}
\end{figure}

A phenomenological example of a screening effect described through Eq.~(\ref{eq:effscr}) is illustrated in the left-hand panel of Fig.~\ref{fig:one}.

\section{Mapping screening mechanisms} \label{sec:mapping}

The parametrisation of the effective gravitational coupling presented in Eq.~(\ref{eq:effscr}) is sufficiently general to incorporate the screening mechanisms encountered in scalar-tensor gravity theories.
This section gives
a few examples of how the different mechanisms can be mapped onto Eq.~(\ref{eq:effscr}).
Thereby, we shall follow the classification of screening effects presented, for instance, in Refs.~\cite{joyce:14,joyce:16}, augmenting it with the linear suppression effects and more generic Einstein gravity limits.
\begin{itemize}
 \item[(i)] Screening at large field values such as in chameleon~\cite{khoury:03a} or symmetron~\cite{hinterbichler:10} models: This screening effect operates in regions where the Newtonian gravitational potential exceeds a given threshold, $|\Psi_{\rm N}|>\Lambda_{\rm T}$. The mapping of this mechanism to Eq.~(\ref{eq:effscr}) will be described in Sec.~\ref{sec:chameleonscr}.
 \item[(ii)] Screening with first derivatives such as in k-mouflage~\cite{babichev:09}: This screening effect is activated when the local gravitational acceleration passes a given threshold value, $|\nabla\Psi_{\rm N}|>\Lambda_{\rm T}^2$. The mechanism can be mapped to Eq.~(\ref{eq:effscr}) as described in Sec.~\ref{sec:kmouflagescr}.
  \item[(iii)] Screening with second derivatives such as in the Vainshtein mechanism~\cite{vainshtein:72}: This screening mechanism operates when curvature or local densities become large, $|\nabla^2\Psi_{\rm N}|>\Lambda_{\rm T}^3$.
  The mapping of this effect onto Eq.~(\ref{eq:effscr}) is presented in Sec.~\ref{sec:vainshteinscr}.
 \item[(iv)] Linear suppression effects such as the Yukawa suppression or linear shielding mechanism~\cite{lombriser:14b}: These effects become important when separations cross the scale set by the linearised mass or sound speed of the field. The mapping to Eq.~(\ref{eq:effscr}) is provided in Sec.~\ref{sec:linscr}.
\end{itemize}
Additionally, by adopting the scaling method of Ref.~\cite{mcmanus:16}, a mapping of general modified gravity theories with an extra scalar degree of freedom that employ a transition from a gravitational modification to an Einstein gravity limit will be examined in Sec.~\ref{sec:einsteinlimits}.
This approach embeds the nonlinear screening effects listed in (i)-(iii) but could also describe mixtures between those or new suppression mechanisms.

\subsection{Screening at large field values} \label{sec:chameleonscr}

First, consider a chameleon model~\cite{khoury:03a} in Brans-Dicke representation with constant Brans-Dicke parameter $\omega>-3/2$ and potential $U=\Lambda+U_0(1-\scal)^{\alpha}$ with $0<\alpha<1$ (see, e.g., Ref.~\cite{lombriser:13c} for a discussion of this model).
For $\scal\sim1$, we get the scalar field equation
\bq
 \nabla^2\scal = -\frac{1}{3+2\omega}\left[ \kappa^2\rhom - \bar{R}_0 \left(\frac{1-\scal}{1-\bar{\scal}_0} \right)^{\alpha-1}\right] \,. \label{eq:chameleonEoM}
\eq
The effective gravitational coupling obtained from solving Eq.~(\ref{eq:chameleonEoM}) for a spherical top-hat overdensity can be described by a thin-shell interpolation~\cite{khoury:03a,li:11b} between the interior and exterior scalar field values with~\cite{lombriser:13b,lombriser:13c}
\bq
 \frac{\Delta\Geff}{G} \approx \frac{1}{3+2\omega} \min \left( 3x -3x^2 + x^3 , 1 \right)
 = \frac{1}{3+2\omega} \left\{1 + \min \left[(x-1)^3 , 0 \right]\right\} \,, \label{eq:Geffchameleon}
\eq
where $x$ denotes the thin-shell thickness and the minimum function accounts for the saturation in the thick-shell regime.
The full solution to more realistic halo density profiles can be found in Ref.~\cite{lombriser:12}.
The dependence of the thin-shell thickness on the physical top-hat radius (here denoted by $r$) may be written schematically as~\cite{lombriser:13c}
\bq
 x \approx -C_1 \, r \left[ \left(1+C_2 r^{-3}\right)^{1/{(\alpha-1)}} - \left(1+C_3 r^{-3}\right)^{1/{(\alpha-1)}} \right] \,, \label{eq:chameleonthsh}
\eq
where $C_i>0$.
$C_1$ is a model, cosmology, and halo mass dependent coefficient~\cite{lombriser:14a}, and $C_2$, $C_3$ ($C_2 \gg C_3$) relate to the interior and exterior scalar field values, respectively. Hence, $\Delta\Geff/G$ has an environmental dependence.

We now cast Eq.~(\ref{eq:chameleonthsh}) into the form of Eq.~(\ref{eq:effscr}).
First, notice that $A=1$ and that we need to describe two suppression effects, for which we shall adopt $N_0=1$ and $N_1=2$ in Eq.~(\ref{eq:effscr}).
From Eq.~(\ref{eq:chameleonthsh}) we find
\bqa
 r \ll C_3^{1/3} \ll C_2^{1/3} & : & x \rightarrow C_1 r \left( \frac{r^3}{C_3} \right)^{1/(1-\alpha)} \,. \label{eq:chameleonin} \\
  r \gg C_2^{1/3} \gg C_3^{1/3} & : & x \rightarrow \frac{C_1 C_2}{1-\alpha} r^{-2} \,, \label{eq:chameleonout}
\eqa
When $x\ll1$ ($x<1/3$) in Eq.~(\ref{eq:chameleonin}), the minimum function in Eq.~(\ref{eq:Geffchameleon}) is governed by the term $3x$ and we therefore have
\bq
 \frac{\Delta\Geff}{G} \approx \frac{3}{3+2\omega} C_1 C_3^{1/(\alpha-1)} r^{(4-\alpha)/(1-\alpha)} \,. \label{eq:chameleonscreened}
\eq
When $x\ll1$ in Eq.~(\ref{eq:chameleonout}) at large scales, the minimum function in Eq.~(\ref{eq:Geffchameleon}) is governed again by the term $3x$ and we therefore have
\bq
 \frac{\Delta\Geff}{G} \approx \frac{3}{3+2\omega} \frac{C_1 C_2}{1-\alpha} r^{-2} \,. \label{eq:Yukawasuppressed}
\eq
Note that the small- and large-scale modifications depend on the exterior and interior densities, respectively.
In order to describe the two suppression effects, we now map these limits onto the parametrisation in Eq.~(\ref{eq:effscr}),
\bq
 \frac{\Delta\Geff}{G} = B \prod_{j=1}^{2} b_j \left(\frac{r}{r_{0j}}\right)^{a_j} \left\{\left[1+\left(\frac{r_{0j}}{r}\right)^{a_j}\right]^{1/b_j} - 1 \right\} \,, \label{eq:effscrchameleon}
\eq
where we drop the indices of the sum in Eq.~(\ref{eq:effscr}) for convenience.
For simplicity, let furthermore $r_{01} \ll r_{02}$, $a_1,b_1>0$, and $a_2<0$, $b_2>0$.

We first consider the scales $r \gg r_{01}$ but where $r \ll r_{02}$ such that contributions from $j=2$ at those scales amount to a factor of unity and only the $j=1$ terms are relevant.
In this regime, we find $\Delta\Geff/G = B$ with the limits described in Eq.~(\ref{eq:limits}) and, thus, $B=(3+2\omega)^{-1}$ from Eq.~(\ref{eq:Geffchameleon}).

In the limit $r \ll r_{01}$, we find
\bq
 \frac{\Delta\Geff}{G} = B \, b_1 \left( \frac{r}{r_{01}} \right)^{a_1(b_1-1)/b_1} \,,
\eq
which can be associated with the chameleon-screened regime, Eq.~(\ref{eq:chameleonscreened}).
Hence,
\bqa
 \frac{B\,b_1}{r_{01}^{a_1(b_1-1)/b_1}} & = & \frac{3}{3+2\omega} C_1 C_3^{1/(\alpha-1)} \, \label{eq:chamrel1} \\
 \frac{a_1}{b_1}(b_1-1) & = & \frac{4-\alpha}{1-\alpha} \,. \label{eq:chamrel2}
\eqa
We need an additional constraint to solve for the $j=1$ parameters in Eq.~(\ref{eq:effscrchameleon}).
However, the above equations are self-adjusting, i.e., if given a $b_1$ then $a_1$ and $r_{01}$ will adjust to satisfy Eqs.~(\ref{eq:chamrel1}) and (\ref{eq:chamrel2}).
Hence, to first approximation, one may simply choose a value for $b_1$.
More accurately, we can introduce an additional constraint from considering the screening scale $r=r_{01}$, where $\Delta\Geff/G = (2^{1/b_1}-1)b_1 B$ from Eq.~(\ref{eq:screeningscaleeq}).
Equating this to Eq.~(\ref{eq:Geffchameleon}) with Eq.~(\ref{eq:chameleonthsh}) evaluated at $r_{01}$
and implementing the relations~(\ref{eq:chamrel1}) and (\ref{eq:chamrel2}) yields an equation for $b_1$ that can be solved numerically.

Let us now consider the Yukawa-suppressed regime with $r \gg r_{01}$ such that at those scales the contributions from $j=1$ amount to a factor of unity and only the $j=2$ terms are relevant.
In the limit $r \ll r_{02}$, we get $\Delta \Geff/G=B$ and in the limit $r \gg r_{02}$, we have
\bq
 \frac{\Delta \Geff}{G} = B\,b_2 \left( \frac{r}{r_{02}} \right)^{a_2(b_2-1)/b_2} \,.
\eq
From comparison to Eq.~(\ref{eq:Yukawasuppressed}), one finds that
\bqa
 \frac{B\,b_2}{r_{02}^{a_2(b_2-1)/b_2}} & = & \frac{3}{3+2\omega}\frac{C_1 C_2}{1-\alpha} \,, \\
 \frac{a_2}{b_2}(b_2-1) & = & -2 \,.
\eqa
We again need an additional constraint to solve for the $j=2$ parameters.
One could choose $b_2$ or as for $j=1$, calibrate the interpolation at $r=r_{02}$
with Eqs.~(\ref{eq:Geffchameleon}) and (\ref{eq:chameleonthsh}) and $\Delta\Geff/G = (2^{1/b_2}-1)b_2 B$ from Eq.~(\ref{eq:screeningscaleeq}).

For completeness, we shall define the $C_i$ functions for the chameleon model studied here~\cite{lombriser:13c}, which with $r\rightarrow\zeta_{\rm h}=a\,\RTH\yhal$ are
\bqa
 C_1 & = & \frac{(3+2\omega)(1-\bscal_0)}{3\Om H_0^2 \RTH^3} \left(\frac{\Om+4\Olam}{4\Olam}\right)^{1/(1-\alpha)} \,, \\
 C_2 & = & \frac{\Om}{4\Olam} \RTH^3 \,, \\\
 C_3 & = & \frac{\Om}{4\Olam} \RTH^3 \left(\frac{y_{\rm h}}{y_{\rm env}}\right)^3 \,.
\eqa
Fig.~\ref{fig:one} illustrates an example of the approximation~(\ref{eq:effscrchameleon}) to the thin-shell description in Eqs.~(\ref{eq:Geffchameleon}) and (\ref{eq:chameleonthsh}) for $\omega=0$ as in $f(R)$ gravity~\cite{buchdahl:70}, $\alpha=1/2$
and $(C_1, C_2, C_3) = (2, 100, 0.1)$, adopting a length scale $\ell$ which is set to unity for simplicity.
Note, however, that $C_3$ is generally a time dependent function.
From the procedure described above, we find
$(a_1,b_1,r_{01})\simeq(8.34,6.21,0.520)$
and
$(a_2,b_2,r_{02})\simeq(-5.82,1.52,28.1)$,
which with $(A,B)=(1,1/3)$ fully describes the interpolation.
Finally, besides chameleon gravity, another model employing screening at large field values is symmetron gravity~\cite{hinterbichler:10}.
Spherical collapse in the symmetron model has, for instance, been studied in Ref.~\cite{taddei:13}.
Thereby, the effective gravitational coupling is given by~\cite{hinterbichler:11}
\bq
 \frac{\Delta\Geff}{G} \approx 6g^2 \left[ x - x^{3/2} \tanh(x^{-1/2}) \right] \,,
\eq
where $g\sim\mathcal{O}(1)$ is related to the symmetry-breaking scalar field value as well as other model parameters, and
schematically
\bq
 x \approx C_1 r \left( 1 - C_2 r^3 \right)^{-1} \,. \label{eq:symmetronthsh}
\eq
With $r\rightarrow\zeta_{\rm h}=a\,\RTH\yhal$, we have $C_1=(\mu^2/\lambda)/(3\Om H_0^2\RTH^3)$ and $C_2=(a\,\RTH)^{-3}$, where $\lambda$ and $\mu^2$ are model parameters defining the scalar field potential.
Note that Eq.~(\ref{eq:symmetronthsh}) is set in a cosmological background.
The symmetron mechanism can be mapped onto the parametrisation in Eq.~(\ref{eq:effscr}) following the same procedure as for chameleon screening.

\subsection{Screening with first derivatives} \label{sec:kmouflagescr}

Next, we consider the k-mouflage mechanism~\cite{babichev:09}, which can operate in scalar-tensor theories with non-canonical kinetic contributions.
Spherical collapse in this model has been studied in Ref.~\cite{brax:14b}.
The scalar field equation in the Einstein frame is given by~\cite{brax:14b}
\bq
 \frac{1}{\sqrt{-\tilde{g}}}\partial_{\mu}\left( \sqrt{-\tilde{g}} \, \partial^{\mu}\phi \, \frac{\rmd K}{\rmd\chi} \right) = \frac{\rmd\ln A}{\rmd\phi} \tilde{\rho}_{\rm m} \,, \label{eq:sfeqkmouflage}
\eq
where $K$ is the non-canonical kinetic term of the scalar field Lagrangian $\mathcal{L}_{\phi}=\mathcal{M}^4K(\chi)$, $A(\phi)=\exp(\beta\kappa\phi)$ defines the conformal factor, relating the Einstein and Jordan frame metrics, $\tilde{g}_{\mu\nu} = A^{-2}(\phi)g_{\mu\nu}$, and $\chi\equiv X/\mathcal{M}^4 \equiv -\partial^{\mu}\phi\partial_{\mu}\phi/(2\mathcal{M}^4)$ with the model parameters $\mathcal{M}$ characterising the suppression scale and $\beta$ the coupling strength.

The effective modification for a static spherically-symmetric matter distribution is given by~\cite{brax:14b,winther:14}
\bq
 \frac{\Delta\Geff}{G} = \frac{2\beta^2}{K_{\chi}(r)} = \frac{\kappa\,\beta}{F_{\rm N}}\sqrt{-2X} \,,
\eq
where the second equality follows from $K_{\chi}^2 X = -2\beta^2 F_{\rm N}^2 /\kappa^2$ with $F_{\rm N}=G\,M(<r)/r^2$ and $M(<r)$ is the mass enclosed within the radius $r$.
We shall adopt the model~\cite{brax:14b}
\bq
 K(\chi) = -1 + \chi + K_0 \chi^2 \,, \ \ \ K_0<0 \,, \label{eq:kmouflagemodel}
\eq
for which we find that schematically
\bq
 \frac{\Delta\Geff}{G} = C_1 \frac{1-\left( x^2 + 1 + x \sqrt{x^2+2} \right)^{1/3}}{x \left( x^2 + 1 + x \sqrt{x^2+2} \right)^{1/6}}
\eq
with $x=C_2 r^{-2}$, $C_2>0$.

For $r\gg\sqrt{C_2}$, i.e., $x\ll1$, we have $\Delta\Geff/G\approx-\sqrt{2}C_1/3$.
For $r\ll\sqrt{C_2}$, i.e., $x\gg1$, we get $\Delta\Geff/G\approx-2^{1/6}C_1 C_2^{-2/3} r^{4/3}$.
Adopting the parametrisation in Eq.~(\ref{eq:effscr}) as approximation with
\bq
 \frac{\Delta G_{eff}}{G} = B \, b \left(\frac{r}{r_0}\right)^a \left\{\left[1+\left(\frac{r_0}{r}\right)^a\right]^{1/b} - 1 \right\}
\eq
and $a,b>0$,
one can infer from $r \gg r_0$ and $r \ll r_0$ with Eq.~(\ref{eq:limits}) that $B = -\sqrt{2}C_1/3$ and $a(b-1)/b = 4/3$, respectively.
Furthermore, $r_0=[2(b/3)^3C_2^2]^{1/4}$ and one may calibrate $b$ at $r=r_0$ using Eq.~(\ref{eq:screeningscaleeq}).

For completeness, we note that
\bqa
 C_1 & = & -3\sqrt{2}\beta^2 \,, \\
 C_2 & = & \frac{3\beta\,\kappa\,M}{2\pi\,\mathcal{M}^2}\sqrt{-3 K_0}
\eqa
for the model adopted in Eq.~(\ref{eq:kmouflagemodel}).

\subsection{Screening with second derivatives} \label{sec:vainshteinscr}

The Vainshtein mechanism~\cite{vainshtein:72} operates, for instance, in DGP braneworld gravity~\cite{dvali:00}, for which the spherical collapse model has been studied in Ref.~\cite{schmidt:09b}.
The brane-bending mode in the subhorizon, quasistatic limit is described by the equation of motion~\cite{koyama:07b}
\bq
 \nabla^2\scal + \frac{r_c^2}{3\beta}\left[ \left(\nabla^2\scal\right)^2 - \left(\nabla_i\nabla_j\scal\right)\left(\nabla^i\nabla^j\scal\right) \right] = \frac{\kappa^2}{3\beta} \drhom \,, \label{eq:dgp}
\eq
where $r_c$ is a crossover scale characterising the impact
of the propagation of the graviton into the codimension of the bulk embedding our 4D brane universe and
\bq
 \beta = 1 + 2\sigma \, H \, r_c \left( 1 + \frac{1}{3}\frac{H'}{H} \right)
\eq
with $\sigma=\pm1$. Positive $\sigma$ represents the normal branch whereas the negative sign is obtained in the self-accelerating branch.
Note that the self-accelerating branch is suffering from a ghost instability~\cite{koyama:07} and the normal branch is strongly constrained by cosmological data~\cite{lombriser:09,raccanelli:12}.
DGP nevertheless serves as useful toy model to study and test modifications of gravity.

The effective gravitational coupling following from Eq.~(\ref{eq:dgp}) for a spherical top-hat matter density profile is schematically given by~\cite{schmidt:09b}
\bq
 \frac{\Delta\Geff}{G} = C_1r^3\left[ \sqrt{1 + C_2 r^{-3}} - 1 \right] \,,
\eq
which can straightforwardly be mapped onto Eq.~(\ref{eq:effscr}).
Hence, we have $A=1$, $N_0=1$, $N_1=1$ with $B=C_1 C_2/2$, $a=3$, $b=2$, and $r_0=C_2^{1/3}$.
For completeness, we note that $C_1 = 2/(3\beta r_v^3)$ and $C_2 = r_v^3$, where we have used the Vainsthein radius $r_v \equiv (2/3)(\kappa^2 \bar{\rho}_{{\rm m}0} r_c^2/\beta^2)^{1/3}\RTH$ for $\rhom\gg\bar{\rho}_{\rm m}$.

The Vainshtein mechanism also operates in galileon gravity~\cite{nicolis:08}, for which the spherical collapse model has been studied in Ref.~\cite{barreira:13}.
The effective gravitational coupling in the cubic model is of the same form as in DGP gravity.
The quartic model introduces weak gravity in the screened regime, which can be modelled in Eq.~(\ref{eq:effscr}) by setting $A<1$.
Note that the quartic model is heavily constrained since also introducing strong deviations in the speed of gravitational waves from the speed of light that cannot be screened~\cite{brax:15}.
Finally, the quintic galileon has no real solution over the full domain~\cite{barreira:13}.
%

\subsection{Yukawa suppression and linear shielding mechanism} \label{sec:linscr}

Lastly, we consider a scalar-tensor theory in Brans-Dicke representation with a potential
$U=-m^2(3+2\omega)\scal/2$
and constant Brans-Dicke parameter $\omega$
such that the 
quasistatic scalar field equation reads
\bq
 \nabla^2\dscal-m^2\dscal+\frac{\kappa^2}{3+2\omega}\drhom \simeq 0 \,.
\eq
For the matter distribution we shall adopt a spherical top-hat overdensity
placed in the cosmological background.
The effective modification of the gravitational coupling may then be written as
\bq
 \frac{\Delta\Geff}{G} = \frac{3x}{3+2\omega}
\eq
with
\bq
 x = \frac{-1 + m\,r + (1 + m\,r) e^{-2m\,r}}{2m ^3 r^3} \,,
\eq
For $r\rightarrow0$, we obtain $\Delta\Geff/G = (3+2\omega)^{-1}$, and in the limit $r\rightarrow\infty$, one finds
\bq
 \frac{\Delta\Geff}{G} = \frac{3}{2(3+2\omega)}\frac{1}{m^2r^2} \,,
\eq

Adopting the parametrisation in Eq.~(\ref{eq:effscr}) with
\bq
 \frac{\Delta G_{eff}}{G} = B \, b \left(\frac{r}{r_0}\right)^a \left\{\left[1+\left(\frac{r_0}{r}\right)^a\right]^{1/b} - 1 \right\}
\eq
and letting $a<0$ and $b>0$, one infers $B = (3+2\omega)^{-1}$ from the limit $r \ll r_0$ and
\bq
 B\,b \left(\frac{r}{r_0}\right)^{a(b-1)/b} = \frac{3}{2(3+2\omega)}\frac{1}{m^2r^2}
\eq
from $r \gg r_0$.
We therefore have $a(b-1)/b = -2$ and
$r_0 = (m\sqrt{2b/3})^{-1}$, where one may again solve for $b$ by calibrating the interpolation at $r_0$ with Eq.~(\ref{eq:screeningscaleeq}).

The Yukawa-like interaction modifies gravity on small scales below the Compton wavelength while it restores GR at large scales, neglecting time derivative terms~\cite{lombriser:15b}.
One can also consider an opposite scenario in which GR is restored on small scales while modifying it on large scales, which is a more natural scenario if aiming at an alternative explanation for cosmic acceleration.
It was shown in Ref.~\cite{lombriser:14b} that scalar-tensor theories can indeed employ a linear shielding mechanism that cancels modifications on small scales while allowing for modifications on large scales.
In Horndeski gravity, the requirement of such a cancellation effect, however, implies a non-standard speed of gravitational waves~\cite{lombriser:15c}.
The parametrisation in Eq.~(\ref{eq:effscr}) is general enough to embed the phenomenology of the linear shielding mechanism.

\section{Mapping general Einstein limits} \label{sec:einsteinlimits}

In addition to the screening mechanisms and linear suppression effects with their mappings to the effective gravitational coupling Eq.~(\ref{eq:effscr}) that enters the spherical collapse equation~(\ref{eq:MGsphcoll}) discussed in Sec.~\ref{sec:mapping},
we next consider a mapping of more generic transitions between modified and Einstein gravity limits that can be encountered in Horndeski scalar-tensor theory~\cite{horndeski:74}.

Sec.~\ref{sec:einsteinlim} gives a brief outline of the scaling method developed in Ref.~\cite{mcmanus:16} and describes how it can be used to identify a recovery of GR for a modified gravity theory.
The method is then adopted in Sec.~\ref{sec:radialprofile} to describe a general procedure for obtaining approximations to general screening effects with Eq.~(\ref{eq:effscr}) from considering the different limits of the effective gravitational coupling, which can then be used in the spherical collapse equation~(\ref{eq:MGsphcoll}).
In Sec.~\ref{sec:examples}, the method is shown to reproduce the radial dependence of the chameleon, k-mouflage, and Vainshtein screening effects as well as the Yukawa suppression described in Sec.~\ref{sec:mapping}.

\subsection{Einstein gravity limits} \label{sec:einsteinlim}

The scaling method was developed in Ref.~\cite{mcmanus:16} to identify the dominant terms in the metric and scalar field equations relevant for performing an expansion in the nonlinear regime of the extra gravitational interaction.
The method has been applied to chameleon gravity and the cubic galileon model to recover the known attributes of the particular screening mechanisms.
More generally, however, it has also been used to derive a set of conditions on the free functions in the Horndeski scalar-tensor action that directly evaluate whether a given embedded model possesses an Einstein gravity limit or not.

For a brief outline, let $\alpha$ denote the coupling associated with the new terms in the gravitational action.
The scaling method then performs an expansion of the scalar field as~\cite{mcmanus:16}
\bq
 \scal=\scal_0(1+\alpha^q\psi) \,, \label{eq:scaling}
\eq
where $|\alpha^q\psi|\ll1$ describes a perturbation in both the limit of $\alpha\rightarrow\infty$ or $0$ when the coupling is relatively strong or weak. 
In order for an Einstein gravity limit to exist, we must recover $R_{\mu\nu}\sim T_{\mu\nu} -g_{\mu\nu}T/2$ as in GR in the metric field equation at leading order with a scalar field equation $D^{(2)}(g_{\mu\nu},\psi)\sim T$, where $D^{(2)}$ contains at most second-order derivatives of the metric and the scalar field.
Moreover, the values for $q$ recovered in these equations of motion and imposed by the particular limits of $\alpha$ have to be consistent among each other, not cause any divergencies in the equations, and balance the contributions of $T$ (see Sec.~\ref{sec:examples} for examples and Ref.~\cite{mcmanus:16} for more details). 

As a first illustration, consider the example scalar field equation
\bq
 \nabla^2\varphi + \alpha (\nabla\varphi)^2 \sim T \,.
\eq
Applying the expansion (\ref{eq:scaling}), one obtains
\bq
 \alpha^q\nabla^2\psi + \alpha^{1+2q} (\nabla\psi)^2 \sim T \,.
\eq
In the limit of $\alpha\rightarrow0$, we must have $q=0$ for no terms to diverge and for one term on the left-hand side to balance the right-hand side.
In this case, the scalar field equation becomes $\nabla^2\psi \sim T$.
In the limit of $\alpha\rightarrow\infty$, we must analogously have $q=-1/2$ and one obtains $(\nabla\psi)^2 \sim T$.
For an Einstein gravity limit to exist when $\alpha\rightarrow\infty$, we would similarly need a recovery of the GR metric field equation at leading order for $q=-1/2$.
We do not delve into this requirement as the focus of this section will be on the radial dependence of $\scal$ and its effect on $\Delta\Geff/G$ (see, however, Ref.~\cite{mcmanus:16} for more details).
We also restrict to spherical symmetries. The extension of the scaling method to symmetry dependent screening effects is discussed in Ref.~\cite{mcmanus:16}.

\subsection{Radial dependencies from a generic scalar field equation} \label{sec:radialprofile}

Consider a generic scalar field equation of the form
\bq
 \sum_i \alpha^{n_i} D_i^{(2)}(g_{\mu\nu},\scal) \sim T \,. \label{eq:genericsfeq}
\eq
We approximate the metric through Minkowski space and apply the expansion of the scalar field in Eq.~(\ref{eq:scaling}).
Taking a formal limit of $\alpha\rightarrow\infty$ or 0,
this will extract the dominant terms in the equation of motion~(\ref{eq:genericsfeq}) such that we arrive at
\bq
 D^{(2)}_{\rm dom}(\psi) \sim T \,.
\eq
We specify to pressureless dust and assume that the system has a symmetry such that only radial derivatives remain.
We then approximate those through
$\nabla \sim r^{-1}$ and
the density through $\rho \sim r^{-v}$, where $v$ depends on the symmetry of the system, typically assumed spherical with $v=3$.
This yields the relation
\bq
 \sum_n a_n(\scal_0,\Delta\scal_0) \left(\frac{\psi}{r^2}\right)^{s_n} \left(\frac{\psi}{r}\right)^{t_n} \psi^{u_n} r^v \sim 1 \,, \label{eq:psiofr}
\eq
where $u_n$ is assumed constant and the exponents $s_n$, $t_n$ denote the powers of second and first derivatives of the scalar field in $D_{\rm dom}^{(2)}(\psi)$, respectively.
Furthermore, $\Delta\scal_0\equiv\scal_n-\scal_0$ with root background values $\scal_n$ and the coefficients $a_n$ are functions of $\scal_0$ and $\Delta\scal_0$.

One can solve Eq.~(\ref{eq:psiofr}) to determine the radial dependence of $\psi$ in the particular limit of $\alpha$, which in the case of a single summand is simply
\bq
 \psi\sim r^{(2s+t-v)/(s+t+u)} \,. \label{eq:psigeneral}
\eq
For the effective gravitational modification in this limit, it follows that
\bq
 \Delta\Geff/G\sim r^{(3s+2t+u-v)/(s+t+u)} \,. \label{eq:Geffgeneral}
\eq
These equations can also be related to the exponent $q$ of $\alpha$ used in the expansion~(\ref{eq:scaling}).
Considering Eq.~(\ref{eq:psiofr}), and assuming the $n_i$ of the dominating term in Eq.~(\ref{eq:genericsfeq}) is unity, which it may be set to if there is, e.g., only one extra term in the action, we find that
\bq
 q=-1/(s+t+u) \,. \label{eq:qgeneral}
\eq
Importantly, note that in Ref.~\cite{mcmanus:16}, it was shown that for Horndeski gravity, $q$ can directly be evaluated from the action of the theory.
It is therefore intriguing to speculate that these relations for the radial dependencies of $\psi$ and $\Delta\Geff/G$ in a specific limit of $\alpha$ could also be directly inferred from the action. This would allow to immediately determine whether the theory in consideration employs a screening mechanism.
For $\Delta\Geff/G$, for instance, one finds the exponent $2-q(s-u-v)$ or $1-q(2s+t-v)$, where $v=3$ for a spherical system.
Sec.~\ref{sec:examples} examines a few examples of these equations from the models discussed in Sec.~\ref{sec:mapping}.
%

\subsection{Examples} \label{sec:examples}

We shall briefly revisit the different suppression mechanisms encountered in Sec.~\ref{sec:mapping} to illustrate the operability of the scaling method.
Note that we will not consider the metric field equations as for the examples given here, these are consistent with the values of $q$ in the corresponding limits of $\alpha$ and recover the Einstein field equations at leading order (see Ref.~\cite{mcmanus:16} for a more complete discussion).
Of interest here are only the radial dependencies of $\scal$ and $\Delta\Geff/G$.

First, consider the chameleon model of Sec.~\ref{sec:chameleonscr}.
Its scalar field equation is of the form
\bq
 \nabla^2\scal - \alpha (1-\scal)^{\tilde{\alpha}-1} \sim -\rho \,.
\eq
Applying the scaling method with $\scal=\scal_0(1+\alpha^q\psi)$, one obtains
\bq
 \scal_0\alpha^q\nabla^2\psi - \alpha (\Delta\scal_0-\scal_0\alpha^q\psi)^{\tilde{\alpha}-1} \sim -\rho \,,
\eq
where $\Delta\scal\equiv(1-\scal)$.
For $\Delta\scal_0\neq0$ and in the limit of $\alpha\rightarrow0$, we must have $q=0$ for no terms to diverge and for one term on the left-hand side of the equation to remain to balance the right-hand side.
Hence, we find $\psi\sim-1/r$ in this limit after applying the above approximations.
For $\Delta\scal_0=0$ and in the limit of $\alpha\rightarrow0$, we have the additional solution $q=(1-\tilde{\alpha})^{-1}$, for which $\psi \sim -r^{3/(1-\tilde{\alpha})}$.
Thus, this yields the limits $\Delta\Geff/G\sim1$ and $\Delta\Geff/G\sim r^{(4-\tilde{\alpha})/(1-\tilde{\alpha})}$, which recovers the radial dependence in the small-scale chameleon-screened regime found in Sec.~\ref{sec:chameleonscr}.

As an example of screening with first derivatives, we have studied the k-mouflage mechanism in Sec.~\ref{sec:kmouflagescr}.
In this case, for the model adopted, the scalar field equation~(\ref{eq:sfeqkmouflage}) can schematically be written as
\bq
 \nabla \left\{ \nabla \scal \left[1+\alpha (\nabla\scal)^2\right] \right\} \sim \rho \,.
\eq
Applying the scaling method, this becomes
\bq
 \scal_0 \nabla \left\{ \nabla \psi \left[\alpha^q + \scal_0^2 \alpha^{1+3q} (\nabla\psi)^2\right] \right\} \sim \rho \,.
\eq
Hence, for no terms to diverge in the limit $\alpha\rightarrow0$ and for the left-hand side to balance the right-hand side of the equation, we must have $q=0$.
With the above approximations, one therefore finds in this limit that $\psi \sim r^{-1}$.
For $\alpha\rightarrow\infty$, we must have $q=-1/3$, which yields $\psi\sim r^{1/3}$.
Thus, one arrives at $\Delta\Geff/G \sim 1$ and $\Delta\Geff/G \sim r^{4/3}$ for $\alpha\rightarrow0$ and $\alpha\rightarrow\infty$, respectively, recovering the results of Sec.~\ref{sec:kmouflagescr}.

In the case of the Vainshtein mechanism, representing screening with second derivatives, we have considered the DGP model in Sec.~\ref{sec:vainshteinscr} with the scalar field equation (\ref{eq:dgp}) schematically given by
\bq
 \nabla^2\scal + \alpha\left[ \left(\nabla^2\scal\right)^2 - \left(\nabla_i\nabla_j\scal\right)\left(\nabla^i\nabla^j\scal\right) \right] \sim \rho \,.
\eq
Applying the scaling method, one obtains
\bq
 \scal_0\alpha^q\nabla^2\psi+\alpha^{1+2q}\scal_0^2 \left[ \left(\nabla^2\psi\right)^2 - \left(\nabla_i\nabla_j\psi\right)\left(\nabla^i\nabla^j\psi\right) \right]
 \sim \rho \,. \label{eq:scaleddgp}
\eq
For the equation not to diverge in the limit of $\alpha\rightarrow\infty$ and for the left-hand side to have one term that balances the right-hand side, we must have $q=-1/2$.
After applying the above approximations, in this limit, we therefore find $\psi\sim\sqrt{r}$.
For $\alpha\rightarrow0$, we must have $q=0$ and thus, $\psi \sim 1/r$.
From this, one infers that $\Delta\Geff/G \sim r^{3/2}$ and $\Delta\Geff/G \sim 1$ in the limits $\alpha\rightarrow\infty$ and $\alpha\rightarrow0$, respectively, which recovers the scaling of the effective modification in Sec.~\ref{sec:vainshteinscr}.
Note that even the interpolation rate $b=2$ can be found if applying the approximation $\nabla \sim r^{-1}$ and $\rho \sim r^{-3}$ to Eq.~(\ref{eq:scaleddgp})~\cite{mcmanus:16}.
This yields $\scal + \alpha\scal^2/r^2 \sim 1/r$ and, schematically, $\Delta\Geff/G\sim r^3(\sqrt{1+r^{-3}}-1)$.

The scalar field equation for the Yukawa suppression described in Sec.~\ref{sec:linscr} is of the form
\bq
 \nabla^2\scal - \alpha \, \dscal \sim -\rho \,,
\eq
where $\alpha=m^2$.
From scaling with $\delta\scal=\scal_0\alpha^q\psi$, we get
\bq
 \scal_0\alpha^q\nabla^2\psi - \scal_0\alpha^{1+q}\psi \sim -\rho \,.
\eq
For $\alpha\rightarrow0$, we need $q=0$ in order for no terms to diverge and thus, $\nabla^2\scal \sim -\rho$.
For $\alpha\rightarrow\infty$, we have $q=-1$ and $\dscal \sim \rho/\alpha$.
With the approximation $\nabla \sim r^{-1}$ and $\rho \sim r^{-3}$, one then finds that $\psi \sim r^{-1}$ and $\psi \sim r^{-3}$ when $\alpha\rightarrow0$ and $\alpha\rightarrow\infty$, respectively, or $\Delta\Geff/G \sim 1$ and $\Delta\Geff/G \sim r^{-2}$.
Hence, this recovers the radial dependencies of the Yukawa suppression found in Sec.~\ref{sec:linscr}.

Finally, note that the radial dependencies and values of $q$ found here agree with the relations (\ref{eq:psigeneral})-(\ref{eq:qgeneral}) inferred from counting powers of second and first derivatives and derivative free terms in the scalar field equations as well as factoring in the radial symmetry imposed on the matter distribution.
This allows an efficient mapping of the radial dependence of $\Delta\Geff/G$ in the screened or suppression limits of a theory to Eq.~(\ref{eq:effscr}), which determines the parameter combination $a(b-1)/b$ with free, calibratable interpolation rate $b$.
The amplitude of the modification $B$ can be found from considering the unscreened limit when $\alpha\rightarrow0$.
The screening scale $r_0$ is generally not directly recovered in this approach but it may be parametrised as described in Sec.~\ref{sec:PPFnl}.
%

\section{A parametrised post-Friedmannian formalism} \label{sec:PPF}

In order to test more systematically the great number of modified gravity and dark sector interaction models proposed~\cite{clifton:11,joyce:14,koyama:15,joyce:16}, we require a generalised description of their implications on the cosmological structure formation.
For this purpose a number of PPF formalisms have been developed addressing the modified evolution of the spatially homogeneous and isotropic background of our Universe and the linear perturbations around it~\cite{uzan:06,caldwell:07,zhang:07,amendola:07,baker:12} (also see Ref.~\cite{gleyzes:14b} for a review on the EFT formalism).
For the nonlinear cosmological structure, interpolation functions have been developed that can be calibrated to $N$-body simulations~\cite{hu:07b,li.y:11}.

Sec.~\ref{sec:PPFlinear} reviews the main concepts behind the linear PPF formalisms and a brief discussion on some extensions to the weakly nonlinear scales is presented in Sec.~\ref{sec:PPFweaklynl}.
In Sec.~\ref{sec:PPFnl}, a PPF formalism for the deeply nonlinear cosmological scales is proposed that is based on the parametrisation of the effective gravitational coupling in the spherical collapse model with Eq.~(\ref{eq:effscr}).
Finally, Sec.~\ref{sec:PPFexamples} describes how the different screening mechanisms discussed in Sec.~\ref{sec:mapping} can be embedded in this new formalism.

\subsection{Cosmological background and linear perturbations} \label{sec:PPFlinear}

The evolution of the spatially homogeneous and isotropic cosmological background in a generalised modified gravity, dark energy, or dark sector interaction model can be described by one free function of time, the scale factor $a(t)$, or equivalently $H(a)$ or the effective equation of state $w(a)$.
PPF frameworks (e.g.,~\cite{uzan:06,caldwell:07,zhang:07,amendola:07,baker:12}) then provide a unified description of the linear perturbations around this background.

Conceptually, we may simply attribute the extra field contributions, modifications of gravity, or interactions to a new effective fluid component $T^{\mu\nu}_{\rm eff}$ that contributes to the conventional Einstein field equations as defined in Eq.~(\ref{eq:Tmunueff}).
Due to the Bianchi identities and energy-momentum conservation of the segregated matter components, energy-momentum of this fluid is separately conserved, $\nabla_{\mu}T^{\mu\nu}_{\rm eff} = 0$.
Thus, we can apply the usual cosmological perturbation theory.
This implies four degrees of freedom in each the perturbation of the metric and the perturbation of the energy-momentum tensor.
Four degrees of freedom are fixed by the Einstein and conservation equations, and another two by a gauge choice.
We then need two closure relations that fix the remaining two degrees of freedom.
These are specified by the particular modified gravity or dark sector model and typically absorb the contributing evolution of the extra fields introduced with the modified theory.
Typically the closure relations are defined by a parametrisation of the modified Poisson equation with an effective gravitational coupling $\mu(a,k)\equiv\Geff/G$ and a parametrisation of the gravitational slip between the metric potentials $\gamma(a,k)\equiv-\Phi/\Psi$ in Fourier space, where both are unity in $\Lambda$CDM, but the parametrisation could also involve two different combinations of $\gamma$ and $\mu$ or characterise different relations between the perturbations.
%

\subsection{Weakly nonlinear scales} \label{sec:PPFweaklynl}

While the simple and generalised treatment of the modified linear perturbations allows for an efficient and consistent computation of the evolution of the cosmological structure from the observable Hubble scales to a few tens of Mpc, it fails at increasingly smaller scales.
The problem becomes even more severe in the presence of screening effects, which complicates tests of gravity in this regime, where, however, there is a great amount of observational data available.

A phenomenological PPF formalism for the description of the nonlinear matter power spectrum in modified gravity theories has, for instance, been proposed in Ref.~\cite{hu:07b} with
\bq
 P(k,z) = \frac{P_{\rm non-GR}(k,z) + c_{\rm nl}\Sigma^2(k,z)P_{\rm GR}}{1+c_{\rm nl}\Sigma^2(k,z)} \,, \label{eq:Pkznl}
\eq
where $P_{\rm non-GR}(k,z)$ is the modified nonlinear matter power spectrum without screening effect and $P_{\rm GR}(k,z)$ is the nonlinear power spectrum in GR with equivalent background expansion history to the modified model.
$\Sigma^2(k,z)$ is a weighting function governing the degree of screening efficiency with the possibly time-dependent $c_{\rm nl}$ controlling the scale of the effect.
Refs.~\cite{hu:07b,koyama:09} proposed the weight
\bq
 \Sigma^2(k,z) = \left[\frac{k^3}{2\pi^2}P_{\rm lin}(k,z) \right]^n \,, \label{eq:Sigmaweight}
\eq
where $P_{\rm lin}(k,z)$ is the linear power spectrum in the modified model.
The exponent $n$ was introduced in Ref.~\cite{koyama:09}, who found that $n=1$ for DGP and $n=1/3$ for $f(R)$ gravity provide good fits to one-loop perturbation computations, which also determine $c_{\rm nl}(z)$.

Eq.~(\ref{eq:Pkznl}) in combination with the weighting defined by Eq.~(\ref{eq:Sigmaweight}) provides a good description of the weakly (or quasi) nonlinear scales of the power spectrum, but increasingly more complicated extensions of the function $\Sigma^2(k,z)$ need to be adopted to correctly describe the more nonlinear scales of $P(k,z)$ measured with $N$-body simulations~\cite{zhao:10b}.
It was shown in Refs.~\cite{lombriser:13c,lombriser:14a} that the spherical collapse model in combination with the halo model, linear perturbation theory, and a simple quasi-nonlinear interpolation motivated by $c_{\rm nl}\Sigma^2(k,z)$ from Eq.~(\ref{eq:Sigmaweight}) and one-loop perturbations, yields an accurate description of the simulated nonlinear matter power spectrum of chameleon $f(R)$ gravity on scales of $k\lesssim10~h\,{\rm Mpc}^{-1}$.
A combination of perturbative computations with one-halo contributions obtained from a generalised modified spherical collapse model therefore seems promising as a nonlinear extension to the linear PPF framework.

Note that a combination of one-loop computations with a simplified one-halo term was performed in Ref.~\cite{brax:13} for variety of scalar-tensor theories, finding a good match to $N$-body simulations.
Progress on a generalisation of the perturbative approach has recently been made in Refs.~\cite{taruya:16,bose:16}.
The focus of this paper, however, shall be on the deeply nonlinear cosmological scales that are not accessible to the perturbative methods.
%

\subsection{A formalism on nonlinear scales} \label{sec:PPFnl}

Next, we want to describe the modifications of gravity at the deeply nonlinear cosmological scales.
For this purpose, we shall propose a parameterisation of the radial dependence of $\Delta\Geff/G$ that enters the spherical collapse calculations.

In Sec.~\ref{sec:radialprofile}, we have observed that the radial dependence of the effective modification of the gravitational coupling represents a particular combination of the powers of second and first derivative terms, $s$ and $t$, respectively, and the derivative free terms $u$ that appear in the scalar field equation as well as the symmetry of the matter distribution characterised by $v$.
More explicitly, we found the radial profile of the effective gravitational coupling $\Delta\Geff/G\sim r^{(3s+2t+u-v)/(s+t+u)}$.
The parameters $s$, $t$, and $u$ generally differ between the limits where the new terms in the action become dominant and when they are subdominant, leading to different radial dependencies in these regimes.
We now need to parametrise the amplitudes of these radially dependent terms.
As we can see from Eq.~(\ref{eq:limits}) they relate to the screening scale $r_0$, for which in Sec.~\ref{sec:mapping}, we have found the dependence on coefficients $C_i\sim\RTH^n$ with some constant exponent $n$.
Hence, these terms generally introduce a mass dependence in $\Delta\Geff/G$.
We also encountered coefficients $C_i\sim(\yenv/\yhal)^n$ that enter through $r_0$ and account for a possible environmental dependence which can arise from the presence of a scalar field potential with a non-vanishing $u$.
In principle, these dependencies could be mixed with $a$ and $b$ in Eq.~(\ref{eq:effscr}) but if specifying $b$ by some constant such that $a$ becomes a constant for constant $u$ then the dependence on mass and environment enters only through $r_0$. Coupling parameters and the background scalar field may enter both through $r_0$ and $B$.
From these observations, we may now construct a parametrisation for the nonlinear modified structure formation described by the spherical collapse model.
We adopt the effective gravitational coupling $\Geff/G$ of Eq.~(\ref{eq:effscr}) introduced in Sec.~\ref{sec:Geff}, for which we consider here a single element, $N_0=N_1=1$ with $A=p_0$.
We first define a constant $b=p_1$ for the interpolation rate parameter.
The maximal unscreened modification shall be defined by $B=(\Delta\Geff/G)_{\rm max}=p_2$, which can be matched to the maximum of the function $\mu$ obtained from the linear PPF formalism in Sec.~\ref{sec:PPFlinear}.
In the screened or suppressed limit, we define $(\Delta\Geff/G)_{\rm scr} \sim r^{p_3}$, where for a given model, $p_3$ can be determined by the scaling method
with Eq.~(\ref{eq:Geffgeneral}), corresponding to the account of derivatives and symmetry in the scalar field equations.
Generally, we replace $r\rightarrow\zeta_{\rm h}=a\,\RTH\,\yhal$ in Eq.~(\ref{eq:effscr}) as in Sec.~\ref{sec:mapping} to perform the spherical collapse calculation in Eq.~(\ref{eq:ydiff}).
Hence, the modification of the gravitational coupling of one element becomes
\bq
 \frac{\Delta\Geff}{G} = B\,b \left( \frac{\yhal}{y_0} \right)^a \left\{ \left[ 1 + \left( \frac{y_0}{\yhal} \right)^a \right]^{1/b} - 1  \right\} \,, \label{eq:nlpar1}
\eq
where
\bq
 a = \frac{p_1}{p_1-1} p_3 \,, \ \ \ \ \
 b = p_1 \,, \ \ \ \ \
 B = p_2 \label{eq:nlpar2}
\eq
and $y_0\equiv r_0/(a\,\RTH)$.
The only quantity left to parametrise is the screening scale $r_0$ or its dimensionless value $y_0$.
It encodes the mass and environmental dependencies of $\Delta\Geff/G$ and, hence, the spherical collapse calculation.
It also includes dependencies on the background scalar field amplitude and the coupling strength.
It is furthermore a function of time, which, however, mixes with the time dependence arising from the background expansion in Eq.~(\ref{eq:ydiff}) and for some modified gravity models, like DGP, from $p_2$.
To include these effects in the modified spherical collapse model of Sec.~\ref{sec:MGsphcoll}, we propose here a simple parametrisation of the form
\bq
 y_0 = p_4 a^{p_5} \left(2G\,H_0 M_{\rm vir}\right)^{p_6} \left(\frac{\yenv}{\yhal}\right)^{p_7} \,. \label{eq:nlpar3}
\eq
If $p_7\neq0$, one can solve a coupled differential equation for $\yenv$, which for an appropriate definition of the environment may typically be assumed to evolve according to Eq.~(\ref{eq:ydiff}) with $\Delta\Geff=0$~\cite{li:12,lombriser:13c}.

In summary, along with a parameter for the effective modification of the gravitational coupling in the fully screened limit, we have introduced seven parameters to describe each transition mechanism that occurs in a modified gravity model.
One parameter can be freely chosen or optimally calibrated, setting the accuracy in the rate of interpolation between the different regimes of the effective coupling.
A second parameter can be matched to linear PPF predictions of this coupling.
A third parameter gives an account of second and first derivatives and derivative free terms that appear in the scalar field equation as well as the radial symmetry of the system.
The remaining four quantities characterise the screening or transition scale: an absolute scale and a parameter each for its time, mass, and environmental dependence.
Note, however, that in some models a more complicated time dependence can also enter through the absolute scale due to a time-dependent maximal effective gravitational coupling.
Examples for how the different screening mechanisms discussed in Sec.~\ref{sec:mapping} map onto the parametrisation defined by Eqs.~(\ref{eq:nlpar1})-(\ref{eq:nlpar3}) are provided in Sec.~\ref{sec:PPFexamples}.

The parametrisation described here allows for the generalised computation of the spherical collapse density $\deltac$ in modified gravity or dark sector interaction models.
One can then use $\deltac$ to describe modified cluster properties such as concentration~\cite{lombriser:13c}, halo bias~\cite{schmidt:08,lombriser:13c}, cluster profiles~\cite{lombriser:11b}, the halo mass function~\cite{schmidt:08,schmidt:09b,li:11b,lombriser:13b,lombriser:13c,barreira:14a,cataneo:16}, or the halo model power spectrum~\cite{schmidt:08,schmidt:09b,lombriser:13c,barreira:14a}.
A similar computation can also be performed to describe modified void properties~\cite{clampitt:12,lam:14}.
As noted in Sec.~\ref{sec:PPFweaklynl}, the one-halo term determined from the use of the cluster profiles and mass functions can also be combined with perturbative methods to improve accuracy at weakly nonlinear scales, which promises to be a useful framework for the nonlinear extension of the PPF formalism.

It should be noted that the parametrisation of $y_0$ in Eq.~(\ref{eq:nlpar3}) with $\Geff$ built by Eq.~(\ref{eq:nlpar1}) allows an exact mapping of all of the approximations discussed for the different screening mechanisms in Sec.~\ref{sec:mapping} (see Sec.~\ref{sec:PPFexamples}). 
As can be seen from the right-hand panel of Fig.~\ref{fig:one}, the $\Geff$ obtained for chameleon gravity very closely follows the thin-shell prediction of Eqs.~(\ref{eq:Geffchameleon}) and (\ref{eq:chameleonthsh}).
The match is even more accurate than the frequently adopted additional approximation of only accounting for the linear term $3x$ in the minimum function of Eq.~(\ref{eq:chameleonthsh}).
Furthermore, as discussed in Sec.~\ref{sec:vainshteinscr}, Eq.~(\ref{eq:effscr}) or equivalently Eqs.~(\ref{eq:nlpar1}) and (\ref{eq:nlpar3}) allow an exact matching of $\Geff$ for the DGP Vainshtein mechanism.
Hence, due to this agreement at the level of the effective gravitational couplings of the models, relevant deviations in the predictions of $\deltac$ are not expected and numerical results for the respective spherical collapse densities are not provided in this paper.

Besides the parametrised interpolation of $P(k,z)$ discussed in Sec.~\ref{sec:PPFweaklynl}, other nonlinear PPF approaches have proposed the phenomenological parametrisation of the variance $\sigma$, the square-root of the integration of the linear matter power spectrum over the Fourier transform of a top-hat window function, to interpolate between the modified and GR regimes, which can be calibrated to $N$-body simulations~\cite{li.y:11}.
This has proven a good approach in modelling the halo mass function and the matter power spectrum in $f(R)$ gravity.
It is worth noting that the relevant quantity in this approach is the peak threshold $\nu\equiv\deltac/\sigma$, where a GR $\deltac$ is adopted and $\sigma$ interpolates between the integrals of a modified and GR linear power spectrum.
Here, we suggest to use $\nu$ with the modifications predicted by the spherical collapse model encoded in $\deltac$ whereas $\sigma$ is computed from a GR power spectrum.
This is due to the extrapolation of the initial densities $\delta_i$ that yield collapse to $\deltac$ with the GR growth function $D_{\rm GR}(a)$ (see Sec.~\ref{sec:MGsphcoll}).
The two formalisms, hence, are comparable and may be mapped (see discussions in Refs.~\cite{lombriser:13b,lombriser:13c}).

In order to avoid the spherical collapse computation and increase the efficiency in observational parameter estimation analyses, one may also use the formalism defined by Eqs.~(\ref{eq:nlpar1})-(\ref{eq:nlpar3}) to elaborate a direct parametrisation of $\deltac$.
Similar approaches have, for instance, been pursued by Refs.~\cite{kopp:13,achitouv:15} to model the modified nonlinear cosmological structure in $f(R)$ gravity or by Ref.~\cite{mead:16} to develop an efficient generalised halo model fit for the matter power spectrum.
Note, for instance, that we recover the scaling of $\Delta\Geff/G\sim(1-\bscal_0)/M_{\rm vir}^{2/3}$ for $f(R)$ gravity~\cite{li.y:11,lombriser:14a} (see Sec.~\ref{sec:PPFexamples}) such that one may rescale a solution of $\deltac$ of one set of parameters to another.
Further relations between the effects of the parameters $p_i$ may allow further rescalings from a fiducial $\deltac$ function.

Besides the application of Eqs.~(\ref{eq:nlpar1})-(\ref{eq:nlpar3}) to the spherical collapse model, one could also use the parametrisation to produce effective $N$-body simulations, using techniques that have been employed to speed-up the simulations~\cite{winther:14} or to rescale from $\Lambda$CDM to modified gravity simulation outputs~\cite{mead:14}.

Finally, note that the deeply nonlinear cosmological scales are affected by baryonic effects, which need to be taken into account when comparing predictions to observations.
These are, however, often modelled with fitting functions that are motivated to directly match observations.
Hence, where lacking a physical prediction, these may conservatively also be adopted to model the gas and stars in the modified scenarios.
Ultimately, with a physical description, baryonic effects may also be used to discriminate between universal and matter-specific couplings on cosmological scales.
Some progress in this direction has been made by including the clustering of baryons in the halo model~\cite{fedeli:14,mead:16} or by imposing self-consistency between the fitting functions of different gas observations, which may account for the effect of modified gravity through its impact on the hydrodynamic equilibrium of the cluster gas~\cite{terukina:13}.
Statistical techniques such a density weighting in the matter power spectrum can also be used to break degeneracies between baryonic and modified gravity effects, or between other cosmological parameters, and may even be used to effectively unscreen the gravitational modifications~\cite{lombriser:15a}.

\subsection{Examples} \label{sec:PPFexamples}

We revisit the screening mechanisms discussed in Sec.~\ref{sec:mapping} to provide a few examples for the nonlinear parametrisation defined by Eqs.~(\ref{eq:nlpar1})-(\ref{eq:nlpar3}).
In all examples $p_0=1$.
For simplicity, we shall specify to collapse today.

For the chameleon model studied in Sec.~\ref{sec:chameleonscr}, in the screened regime, we obtain for a choice of $p_1$,
\bq
 \begin{array}{lll}
  p_2 = \frac{1}{3+2\omega} \,, \ \ \ &  p_3 = \frac{4-\alpha}{1-\alpha} \,, \ \ \ & p_4 = \Om^{1/3}\left[(\Om + 4\Omega_{\Lambda})^{1/(\alpha-1)} \frac{p_1 p_2}{1-\bscal_0} \right]^{1/p_3} \,, \\
  p_5 = -1 \,, & p_6 = \frac{2}{3 p_3} \,, & p_7 = \frac{3}{\alpha-4} \,.
  \label{eq:chameleonppf}
 \end{array}
\eq
The Yukawa-suppressed regime, given $p_1$, is described by
\bq
 \begin{array}{lll}
  p_2 = \frac{1}{3+2\omega} \,, \ \ \ &  p_3 = -2 \,, \ \ \ &
  p_4 = \Om^{1/3} \left\{  (1-\alpha) \left[  (4\Omega_{\Lambda})^{\alpha-2}(\Om + 4\Omega_{\Lambda}) \right]^{1/(\alpha-1)} \frac{p_1p_2}{1-\bscal_0} \right\} ^{1/p_3} \,, \\
  p_5 =  -1 \,, & p_6 = \frac{2}{3p_3}  \,, & p_7 = 0 \,.
  \label{eq:yukawappf}
 \end{array}
\eq
Note that we have a mass dependence in both relations and the chameleon regime, Eq.~(\ref{eq:chameleonppf}), is also environment dependent.

The k-mouflage model studied in Sec.~\ref{sec:kmouflagescr} is described by
\bq
 \begin{array}{lll}
  p_2 = 2\beta^2 \,, \ \ \ &  p_3 = \frac{4}{3} \,, \ \ \ & p_4 = \Om^{1/3} \left[p_1 \left(- 4 p_2 K_0\right)^{1/3} \right]^{1/p_3} \left(\frac{H_0}{\kappa\,\mathcal{M}^2} \right)^{1/2} \,, \\
  p_5 =  -1 \,, & p_6 = \frac{1}{3}\left(\frac{2}{p_3}-1\right) \,, & p_7 = 0 \,,
  \label{eq:kmouflageppf}
 \end{array}
\eq
which is mass but not environment dependent.

For the Vainshtein screening of DGP studied in Sec.~\ref{sec:vainshteinscr}, we get $p_1=2$ and
\bq
 \begin{array}{lll}
  p_2 = \frac{1}{3\beta} \,, \ \ \ &  p_3 = \frac{3}{2} \,, \ \ \ &
  p_4 = 2 \Om^{1/3} p_2^{2/3} (H_0 r_c)^{2/3} \,, \\
  p_5 =  -1 \,, & p_6 = 0  \,, & p_7 = 0 \,.
  \label{eq:vainshteinppf}
 \end{array}
\eq
Hence, we have no mass or environmental dependence but a radial or density dependence still enters the spherical collapse computations through $\yhal$ in Eq.~(\ref{eq:nlpar1}).

Finally, the Yukawa-suppressed scenario in Sec.~\ref{sec:linscr} is recovered for
\bq
 \begin{array}{lll}
  p_2 = \frac{1}{3+2\omega} \,, \ \ \ &  p_3 = -2 \,, \ \ \ & p_4 = \Om^{1/3} \left(\frac{2p_1}{3}\right)^{1/p_3} \frac{H_0}{m} \,, \\
  p_5 =  -1 \,, & p_6 = \frac{2}{3p_3}  \,, & p_7 = 0 \,.
  \label{eq:linscrppf}
 \end{array}
\eq
Note that the $p_i$ are dimensionless and the units left in some of the $p_4$ cancel out.

\section{Conclusions} \label{sec:conclusions}

This paper proposes a parametrisation of the modified gravity effects that can manifest in the nonlinear cosmological structure and may be used to test gravity with the wealth of observations available in this regime.
The formalism is based on a generalisation of the modified gravitational forces that enter the spherical collapse model calculations and embeds the variety of screening mechanisms available to Horndeski scalar-tensor theories.
A discussion of how more generic transitions between modified and Einstein gravity limits can be mapped onto the parametrised effective gravitational coupling is also presented.
The results are then used to formulate a nonlinear extension to the linear parametrised post-Friedmannian formalism that enables the generalised examination of modified gravity with the deeply nonlinear cosmological structure of dark matter which is amenable to spherical collapse model predictions.
It may further be combined with perturbative methods covering the quasi-nonlinear regime and baryonic modelling.
A first parameter in this formalism describes a deviation from the Newtonian gravitational constant in the fully screened limit, which can be neglected if a theory recovers GR.
Along with a screening scale, each transition then includes up to three parameters:
an amplitude for the modification of the gravitational coupling in the fully unscreened limit, the exponent of a power-law radial profile of the screened solution, and an interpolation rate between the screened and unscreened limits.
The amplitude of the unscreened modification can be matched to a linear PPF prediction, and the power of the screened radial profile represents an account of second and first derivatives and derivative free terms that appear in the scalar field equation as well as the radial symmetry of the system.
Each screening scale can then generally be modelled introducing an absolute dimensionless scale and a parameter each for its time, mass, and environmental dependence.

This simple framework enables the generalised computation of the spherical collapse density in modified gravity or dark sector interaction models, embedding the known screening mechanisms of viable second-order scalar-tensor theories, which can then be used to describe modified cluster properties, the halo model power spectrum, or may be altered to describe modified void properties.
The results could also be used to elaborate a direct parametrisation of the spherical collapse density, for instance, by taking advantage of scaling relations that can be identified in the parameterisation of the effective gravitational coupling and may allow the extrapolation from a fiducial spherical collapse density. 
Such an approach would further increase the efficiency in predicting modified gravity effects in the nonlinear cosmological structure, suitable for an implementation in parameter estimation analyses when testing gravity.
Finally, the parametrisation may also be used with techniques that have been employed to speed-up $N$-body simulations or to rescale from $\Lambda$CDM to generalised modified gravity simulation outputs.
The examination of such applications is left for future work.
%

\acknowledgments

The author thanks Matteo Cataneo, Ryan McManus, Francesco Pace, and Andy Taylor for useful discussions and comments on the manuscript.
This work was supported by a SNSF Advanced Postdoc.Mobility Fellowship (No.~161058).
The author also acknowledges support from the STFC Consolidated Grant for Astronomy and Astrophysics at the University of Edinburgh.
Access to research materials can be provided on demand.

\newpage
\bibliographystyle{JHEP} 
\bibliography{bib}

\end{document}